\documentclass[letterpaper, 10 pt, conference]{ieeeconf}  

\IEEEoverridecommandlockouts                             

\usepackage{graphics} 
\usepackage{epsfig} 
\usepackage{mathptmx} 
\usepackage{times} 
\usepackage{amsmath} 
\usepackage{amssymb}  
\usepackage{xcolor}
\usepackage{mathtools}
\usepackage{gensymb}
\usepackage{arydshln}
\usepackage{colortbl}

\graphicspath{{./figures/}}

\title{\LARGE \bf
An Improved Height Difference Based Model of Height Profile for Drop-on-Demand 3D Printing With UV Curable Ink
}

\author{Yumeng Wu$^{1}$ and George Chiu$^{2}$
\thanks{*This work was not supported by any organization}
\thanks{$^{1}$Yumeng Wu is with the School of Mechanical Engineering, College of Engineering, Purdue University
585 Purdue Mall, West Lafayette, IN 47907, USA
        {\tt\small wu350@purdue.edu}}%
\thanks{$^{2}$George Chiu is with the School of Mechanical Engineering, College of Engineering, Purdue University
585 Purdue Mall, West Lafayette, IN 47907, USA
        {\tt\small gchiu@purdue.edu}}%
}

\begin{document}

\maketitle
\thispagestyle{empty}
\pagestyle{empty}

\begin{abstract}
        This paper proposes an improved height profile model
        for drop-on-demand 3D printing with UV curable ink.
        It is extended from a previously validated model and computes
        height profile indirectly from volume and area propagation to ensure volume conservation.
        To accommodate 2D patterns using multiple passes,
        volume change and area change within region of interest are modeled as a piecewise function of height difference before drop deposition.
        Model coefficients are experimentally obtained and validated with bootstrapping of experimental samples.
        Six different drop patterns are experimentally validated.
        The RMS height profile errors for 2D patterns from the proposed model are consistently smaller than existing
        models from literature and are on the same level as 1D patterns reported in our previous publication.
\end{abstract}

\section{INTRODUCTION}

Additive manufacturing (AM) or 3D printing,
reduces material wastes and costs to change,
in comparison to the traditional computer numerical control (CNC) machining \cite{ford2016additive}.
This makes small to median batch production economically viable.
As a result,
many industries,
such as aerospace, automotive and biomedical will be benefited from it in manufacturing,
in addition to traditional application as prototyping \cite{wohlers2013wohler}\cite{kupper2017get}.

Among different additive manufacturing processes,
drop-on-demand 3D printing
shares similar advantages,
such as wide range of material, lower cost and as inkjet printer higher resolution than other 3D printing processes \cite{cooley2002applicatons}\cite{sirringhaus2003inkjet}.
There have been many studies focus on developing functional materials specifically for drop-on-demand 3D printing,
especially in biomedical and pharmaceutical area \cite{gantenbein2018three}\cite{Guo_Patanwala_Bognet_Ma_2017}\cite{Cheng_Zheng_Wang_Sun_Lin_2020}\cite{He_Foralosso_Trindade_Ilchev_Ruiz_2020}.
Combined with its ability to fulfill  the demands on both geometry and functions
drop-on-demand 3D printing is more common with biomedical applications \cite{simpson2016preparing}.
UV inkjet printing served as a platform to produce solid oral dosage forms ink specifically developed for ropinirole HCl, a low dose water soluble drug \cite{Clark_Alexander_Irvine_Roberts_Wallace_Sharpe_Yoo_Hague_Tuck_Wildman_2017}
Patients can receive customized endovascular aneurysm repair (EVAR) with 3D printing technology \cite{lei2020new}.
Personalized hearing aids fit patients' ear profiles better with additive manufacturing \cite{chu2008design}.

On the other side,
there is limited work focusing on meeting the geometry demand.
A single drop is usually modeled as a spherical cap after solidification \cite{Doumanidis2000}.
For drops deposited on uneven surface,
drop profile changes and cannot be easily obtained.
Due to material flow and other factors.
Some researchers use computational methods to model the profile \cite{gunjal2003experimental}\cite{Xu_Basaran_2007}\cite{choi2017numerical}.
The estimations are generally close to the real drops,
however,
it takes significant computation time. 
As a result,
it is not viable to be used for real time process or process control.
Other researchers simplify the model to balance between accuracy and computation. 
The graph-based model \cite{Guo2018} was proposed to capture the dynamics of height propagation.
However, it does not guarantee volume conservation and the reported error is often greater than 10\%.
A model based on volume and area propagation based on height difference reduces error in height prediction,
but it is limited to 1D patterns, such as a line \cite{WuY.2019Mhpf}\cite{WuY.2020Mhpf}.

In this paper,
an expanded model is proposed to achieve same level of performance as in \cite{WuY.2020Mhpf} but for more general 2D patterns.
The height profile is obtained indirectly from volume and area propagation to ensure volume conservation.
Both change of volume and change of area within the region of interest of a drop are modeled as a piecewise function of prior height.
Coefficients \(m_v^+\), \(m_v^-\), \(m_a^+\) and \(m_a^-\)  are introduced to quantify the impact of the prior height difference on the drop distribution.
The remaining of the paper is organized as follow.
The model is introduced in Section \ref{sec:model_description}.
Experimental setup is presented in Section \ref{sec:exp_setup}.
Coefficients are obtained in Section \ref{sec:coeff}.
Experimental validation is included in Section \ref{sec:results}.
Lastly, Section \ref{sec:conclusion} is the conclusion.

\section{\MakeUppercase{Model Description}}
\label{sec:model_description}

\begin{figure}
        \centering
        \includegraphics[width=\linewidth]{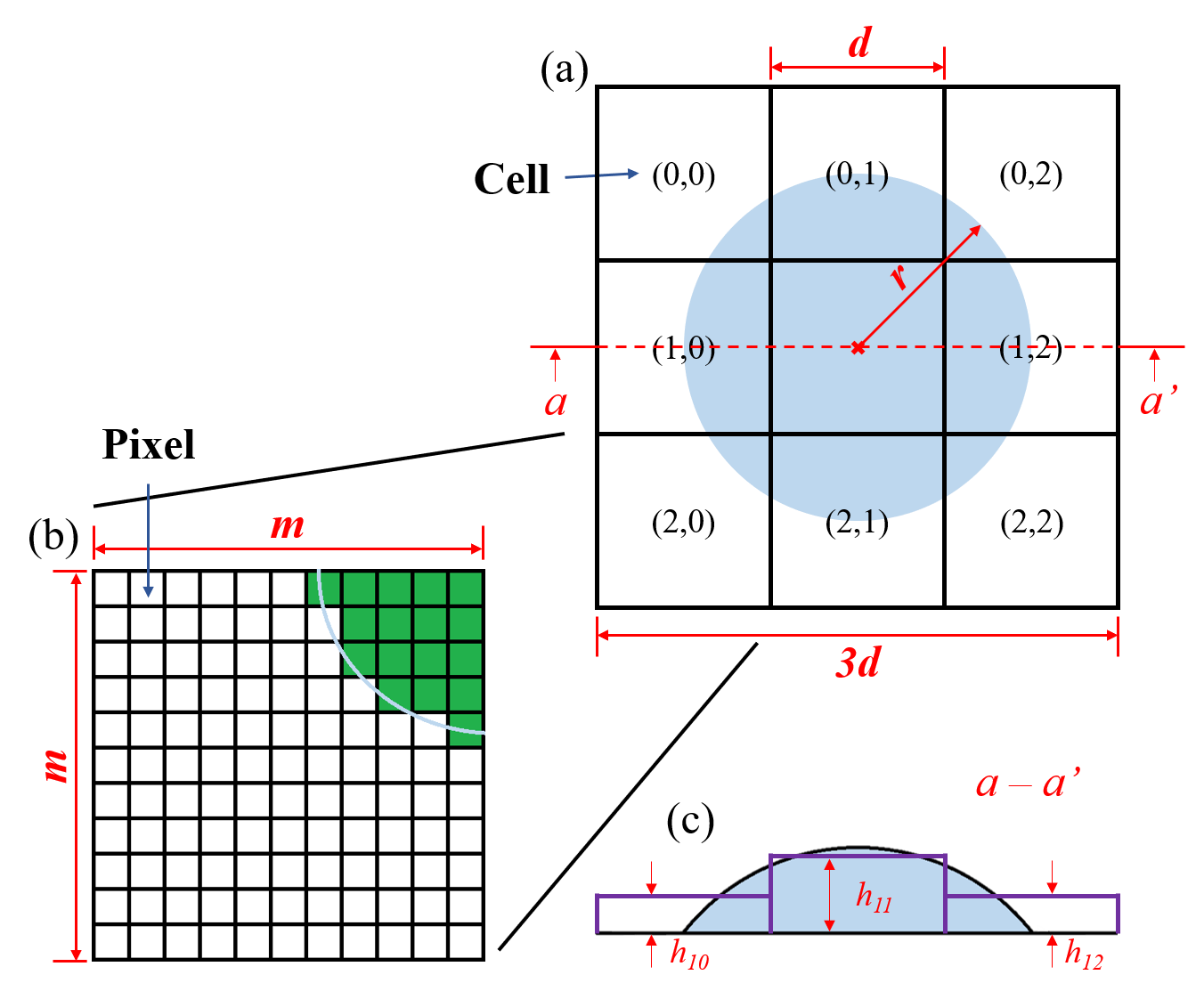}
        \caption{(a): Single drop with radius $r$ on $3 \times 3$ cell. \hspace{\textwidth}
                (b): Zoomed-in section of cell (2,0) in (a), illustrating how to count area covered by inks.
                (c): Height profile comparison of center row. Proposed height profile is in purple, while spherical cap height profile is in light blue.}
        \label{fig:single}
\end{figure}

In \cite{WuY.2020Mhpf},
the height profile of ink-jet printing of UV curable ink is modeled for printing 1D pattern, such as a line.
The dynamic model is validated with experimental data.
However, its height difference is only calculated across each column,
which limited the model's application to general 2D patterns.

The single drop profile is modeled as a spherical cap, the same as that in \cite{WuY.2020Mhpf}.
To accommodate the printing of more than 1 row,
the notations are adjusted accordingly.

After the ink is deposited at the target location,
the radius of the cured drop on the substrate is \(r\),
which is a constant once the printing parameters are set.
The pitch distance \((d)\) is assumed to be known and
ensures sufficient overlap between adjacent drops.
In this article,
\(d\) is chosen to be between \(\frac{r}{2}\) and \(\frac{2r}{3}\).
A cell system is used as the coordinate.
The size of each cell is \(d\times d\).
When a drop is deposited at the center of cell \((i,j)\),
it has impact over \(3 \times 3\) surrounding cells.
Such impact is defined as the region of interest (ROI) of cell \((i,j)\), denoted as \(B_{i j}\).
Mathematically, it can be expressed as
\begin{equation}
        \label{eq:bk}
        B_{i j} =  \{(p,q)|i \in \{p-1,p,p+1\}, j \in \{q-1,q,q+1 \} \},
\end{equation}
where \(i\) and \(j\) represent the row and column index, respectively.
Figure~\ref{fig:single}.(a) shows an example of \(B_{i j}\), where \((i,j)\) = (1,1).
The light blue disk represents the drop.

The percent volume of a single drop, the percent area covered by the ink and the height in cell \((i,j)\) after the \(k^{th}\) drop are denoted as \(v_{ij}[k]\), \(a_{ij}[k]\) and \(h_{ij}[k]\), respectively.
The height is the ratio between volume and area, which can be written as
\begin{equation}
        \label{eq:h_v_k}
        h_{ij}[k] = c \frac{v_{ij}[k]}{a_{ij}[k]},
\end{equation}
where c is a scaling factor that converts the relative number to absolute height and is dependent on equipment and printing parameters.
The height matrix, \(H_{i j}[k]\), represents the cell height within the region of interest, \(B_{i j}\),
after a drop associated with the \(k^{th}\) pass is deposited in cell \((i,j)\).
Mathematically, it can be written as
\begin{equation}
        \label{eq:h_k}
        H_{i j}[k] = \left[ \begin{array}{ccc}
                        h_{i-1 j-1}[k] & h_{i-1 j}[k] & h_{i-1 j+1}[k] \\
                        h_{i j-1}[k]   & h_{i j}[k]   & h_{i j+1}[k]   \\
                        h_{i+1 j-1}[k] & h_{i+1 j}[k] & h_{i+1 j+1}[k]
                \end{array} \right],
\end{equation}
where \(h_{i j}[k]\) represents the height in cell \((i,j)\).
Similarly, the volume matrix \((V_{i j}[k])\) can be written as
\begin{equation}
        \label{eq:v_k}
        V_{i j}[k] = \left[ \begin{array}{ccc}
                        v_{i-1 j-1}[k] & v_{i-1 j}[k] & v_{i-1 j+1}[k] \\
                        v_{i j-1}[k]   & v_{i j}[k]   & v_{i j+1}[k]   \\
                        v_{i+1 j-1}[k] & v_{i+1 j}[k] & v_{i+1 j+1}[k]
                \end{array} \right],
\end{equation}
where \(v_{i j}[k]\) represents the percent volume of a single drop in cell \((i,j)\).
The area matrix, \(A_{i j}[k]\), can be written as
\begin{equation}
        \label{eq:a_k}
        A_{i j}[k] = \left[ \begin{array}{ccc}
                        a_{i-1 j-1}[k] & a_{i-1 j}[k] & a_{i-1 j+1}[k] \\
                        a_{i j-1}[k]   & a_{i j}[k]   & a_{i j+1}[k]   \\
                        a_{i+1 j-1}[k] & a_{i+1 j}[k] & a_{i+1 j+1}[k]
                \end{array} \right],
\end{equation}
where \(a_{i j}[k]\) represents the percent area in cell \((i,j)\).
In addition,
we introduce a height difference matrix \(\tilde{H}_{i j}[k]\) to represent the height difference between adjacent cells within \(B_{i j}\),
which can be written as
\begin{equation}
        \label{eq:h_d}
        \tilde{H}_{i j}[k] = \left[ \begin{array}{ccc}
                        \tilde{h}_{i-1 j-1}[k] & \tilde{h}_{i-1 j}[k] & \tilde{h}_{i-1 j+1}[k] \\
                        \tilde{h}_{i j-1}[k]   & \tilde{h}_{i j}[k]   & \tilde{h}_{i j+1}[k]   \\
                        \tilde{h}_{i+1 j-1}[k] & \tilde{h}_{i+1 j}[k] & \tilde{h}_{i+1 j+1}[k]
                \end{array} \right],
\end{equation}
where \(\tilde{h}_{i j}[k]\) is the height difference between adjacent cells within \(B_{i j}\),
which can be written as
\begin{equation}
        \label{eq:h_d_s}
        \tilde{h}_{i j}[k] = h_{ij}[k] - \overline{h_{p,q}[k]},
\end{equation}
where \((p,q) \in B_{i j}, |p-i| \le 1, |q-j| \le 1\) and \( (p,q) \neq (i,j)\).
For example,
\begin{equation}
        \label{eq:hd_eg0}
        \begin{aligned}
                \tilde{h}_{ij}[k] = & h_{ij}[k]  -\frac{1}{8}(h_{i-1j-1}[k]+h_{i-1j}[k] \\
                                    & +h_{i-1j+1}[k]+h_{ij-1}[k]+h_{ij+1}[k]            \\
                                    & +h_{i+1j-1}[k]+h_{i+1j}[k]+h_{i+1j+1}[k])
        \end{aligned},
\end{equation}
and
\begin{equation}
        \label{eq:hd_eg1}
        \begin{aligned}
                \tilde{h}_{ij-1}[k] = & h_{ij-1}[k]-\frac{1}{5}(h_{i-1j-1}[k] \\
                                      & +h_{i-1j}[k]+h_{ij}[k]                \\
                                      & +h_{i+1j-1}[k]+h_{i+1j}[k])
        \end{aligned}.
\end{equation}

Following our previous work,
individual cell height can be written as
\begin{equation}
        \label{eq:hk}
        h_{i j}[k] = c \frac{v_{i j}[k-1]+\Delta v_{i j}[k]}{a_{i j}[k-1]+\Delta a_{i j}[k]},
\end{equation}
where \(\Delta v_{i j}[k]\) and \(\Delta a_{i j}[k]\) are the percent
volume change in cell \((i,j)\) associated with a drop on cell \((i,j)\) on
the \(k^{th}\) pass and percent area change in cell \((i,j)\)
associated with a drop on cell \((i,j)\) on the \(k^{th}\) pass, respectively.

\subsection{Propagation Model}
When the substrate is flat,
each drop profile is the same and
a simple superposition model would be suffice.
When the substrate is not flat prior to the deposition,
the ink flows from higher to lower within a small range.
To capture this effect,
we introduce a model based on the height difference among adjacent cells.
The height difference at higher cells will be positive and
that at lower cells will be negative.
The percentage volume change due to the drop deposited in cell \((i,j)\) at the \(k^{th}\) pass
\((\Delta v_{ij}[k])\) is a function of the corresponding height difference \((\tilde{h}_{ij}[k])\)
in cells other than the deposition location.
This relationship can be written as
\begin{equation}
        \label{eq:dvk_func}
        \begin{aligned}
                (p,q) \in B_{ij}    & \hspace{1em}\text{and} \hspace{1em} (p,q) \neq (i,j) \\
                \Delta v_{p q}[k] = & v_{p q}[k] - v_{p q}[k-1]                            \\
                =                   & f(\tilde{h}_{pq}[k])                                 \\
                =                   & \left\{
                \begin{array}{ll}
                        m_v^+ \tilde{h}_{pq}[k] & \tilde{h}_{pq}[k]>0 \\
                        m_v^- \tilde{h}_{pq}[k] & \tilde{h}_{pq}[k]<0
                \end{array}
                \right. ,
        \end{aligned}
\end{equation}
where \(m_v+\) and \(m_v^-\) are the percent volume change corresponding a positive  and negative \(\tilde{h}_{ij}[k]\), respectively.
\(m_v+\) and \(m_v^-\) are determined empirically from the first two drops.
For the cell at deposition location \((i,j)\),
\(\Delta v_{ij}[k]\)  is the remaining volume of a drop of ink,
which can be written as
\begin{equation}
        \label{eq:dvk_ij}
        \Delta v_{i j}[k] = 1 - \sum_{p=i-1}^{i+1}\sum_{q=j-1}^{j+1} \Delta v_{p q}[k] \hspace{1em} (p,q) \neq (i,j).
\end{equation}
Thus,
the total volume change within \(B_{i j}\) is guaranteed to be 1, i.e. volume conservation is ensured.

Similarly,
the relationship can be extended to the cell area change
due to the drop deposited in cell \((i,j)\) at the \(k^{th}\) pass \(\Delta a_{ij}[k]\).
However,
the upper limit of a cell area is 1, when it is fully occupied.
As a result,
area modeling is only necessary when the cell has not been fully occupied and it is not the deposition location.
Mathematically,
this can be written as
\begin{equation}
        \label{eq:dak_func}
        \begin{aligned}
                (p,q) \in B_{ij}   &                                                    \\
                \Delta a_{pq}[k] = & a_{p q}[k] - a_{p q}[k-1]                          \\
                =                  & g(\tilde{h}_{pq}[k])                               \\
                =                  & \left\{
                \begin{array}{ll}
                        m_a^+ \tilde{h}_{pq}[k] & \tilde{h}_{pq}[k]>0 \\
                        m_a^- \tilde{h}_{pq}[k] & \tilde{h}_{pq}[k]<0
                \end{array}\right.                                       \\
                \text{and} \hspace{1em}
                a_{p q}[k] =       & \min {\left(a_{p q}[k-1]+\Delta a_{p q},1\right)},
        \end{aligned}
\end{equation}
where \(m_a^+\) and \(m_a^-\) are the area change corresponding a positive  and negative \(\tilde{h}_{ij}[k]\), respectively.
\(m_a^+\) and \(m_a^-\) are determined empirically from the first two drops as with \(m_v^+\) and \(m_v^-\).

\section{EXPERIMENTAL SETUP}
\label{sec:exp_setup}

\begin{figure}
        \vspace{1ex}
        \centering
        \includegraphics[width=0.75\linewidth]{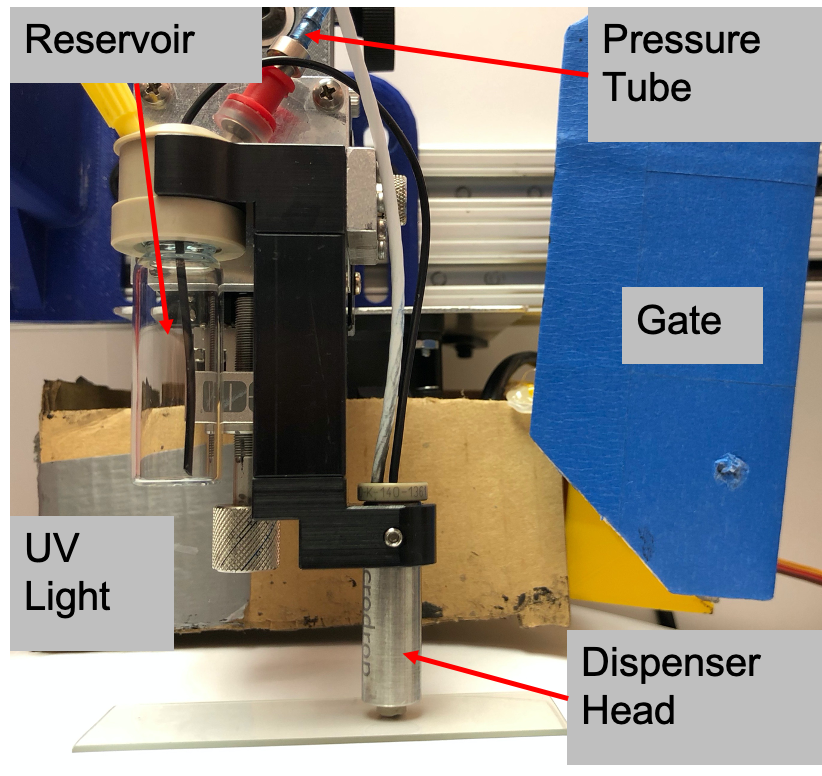}
        \caption{The experimental setup. Gate closes when UV light is on to protect the dispenser head.}
        \label{fig:experiment}
\end{figure}

The experimental setup is shown in Figure~\ref{fig:experiment}.
The experimental equipment includes a Microdrop piezo-electrical dispensing system with a heated 70 \(\mu m\) nozzle,
a Zeta-20 optical profilometer with a 50x objective lens and a 0.35x coupler,
a PI Precision XY stage,
a UV light, a gate and UV inks from Kao Collins Inc.
A dedicate program controls the entire printing process,
depositing each drop at the desired location, curing inks after each layer
and protecting the dispenser head by closing the gate when UV curing is on.
The Z-resolution of the profilometer at this setting is 0.04 $\mu m$ and the pixel area $(a_c)$ is 0.49 $\mu m^2$

The volume of a single drop \((v_s)\) is assumed to be
constant with the same printing parameters.
\mbox{Figure \ref{fig:single}.(b)} is the zoomed-in view of the cell \((2,0)\) in Figure~\ref{fig:single}.(a),
illustrating the measurements taken from the optical profilometer.
There are \(m \times m\) pixels for each cell.
The area of each pixel, \(a_c\) is \(d^2/m^2\).
Each pixel has a height measurement \(h_{i_m,j_m}\) and is marked green if it is covered by the ink.
\(a_{ij}[k]\) is obtained by counting the number of green pixels.
If there are  \(M\) green pixels,
\(a_{ij}[k]\) can be written as
\begin{equation}
        \label{eq:cell_area_pct}
        a_{ij}[k] = \frac{M}{m^2} \times 100\%.
\end{equation}
\(v_{ij}[k]\) is computed by
\begin{equation}
        \centering
        \label{eq:single}
        v_{i j}[k] = \frac{\sum_{i_m=0}^{m-1}\sum_{j_m=0}^{m-1} h_{i_m j_m} a_c}{v_s} \times 100\%.
\end{equation}
\(h_{ij}[k]\) can be obtained by
\begin{equation}
        \centering
        \label{eq:h_single}
        \begin{split}
                h_{i j}[k] &= c \frac{v_{i j}[k]}{a_{i j}[k]} \\
                &= \frac{\sum_{i_m=0}^{m-1}\sum_{j_m=0}^{m-1} h_{i_m j_m}}{M}, \\ \text{and \hspace{0.1in} }
                c & = \frac{v_s}{d^2},
        \end{split}
\end{equation}
where the scaling factor \(c\) is 7.0751
for the experimental setup and the ink used.

\section{OBTAIN \(m_v^+, m_v^-, m_a^+\) AND \(m_a^-\)}
\label{sec:coeff}

\begin{figure}
        \includegraphics[height=12em]{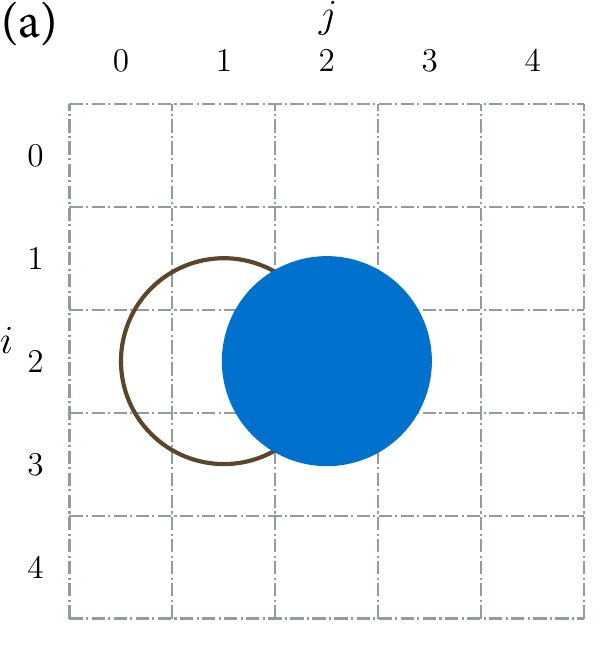} \hspace{1em}
        \includegraphics[height=12em]{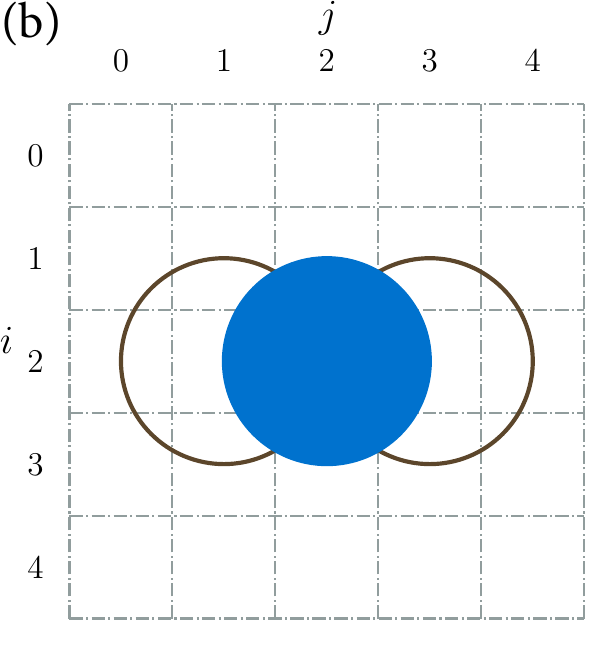}\vspace{1em}\\
        \includegraphics[height=12em]{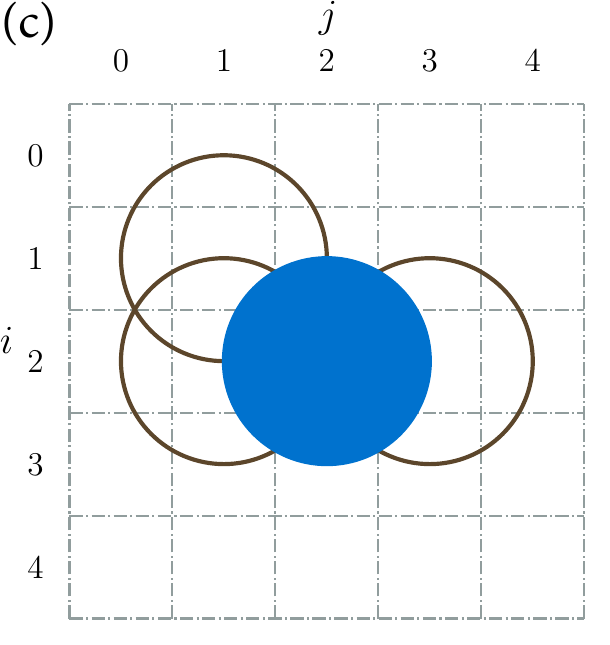} \hspace{1em}
        \includegraphics[height=12em]{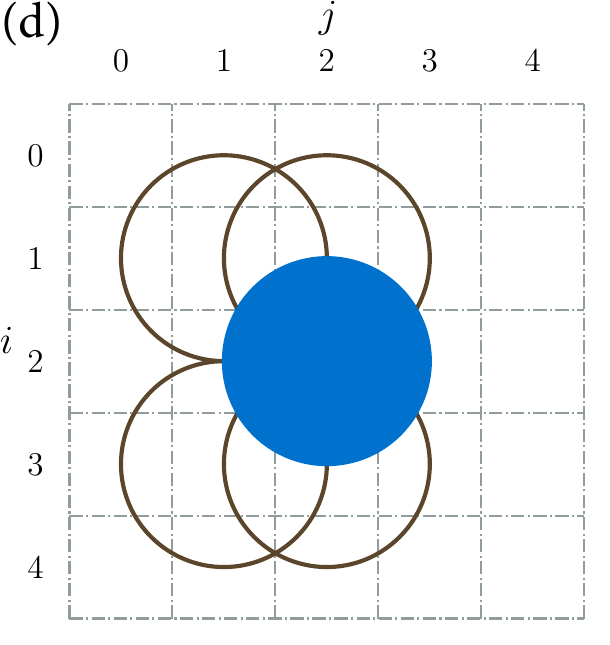}\vspace{1em}\\
        \includegraphics[height=12em]{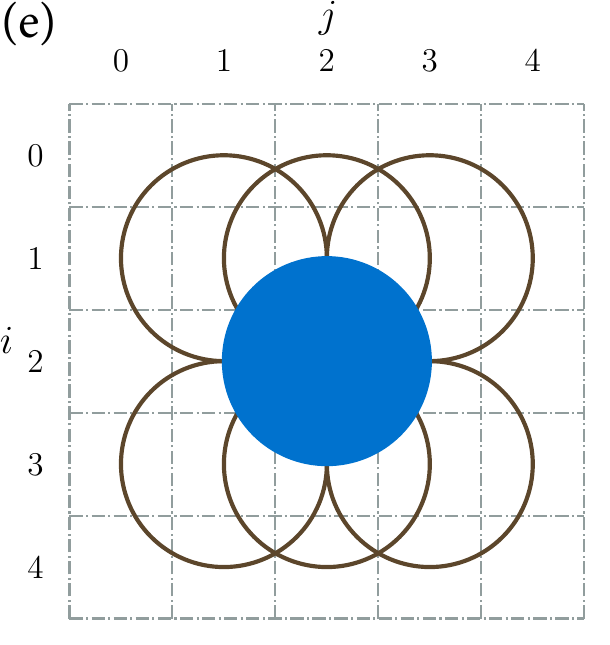}\hspace{1em}
        \includegraphics[height=12em]{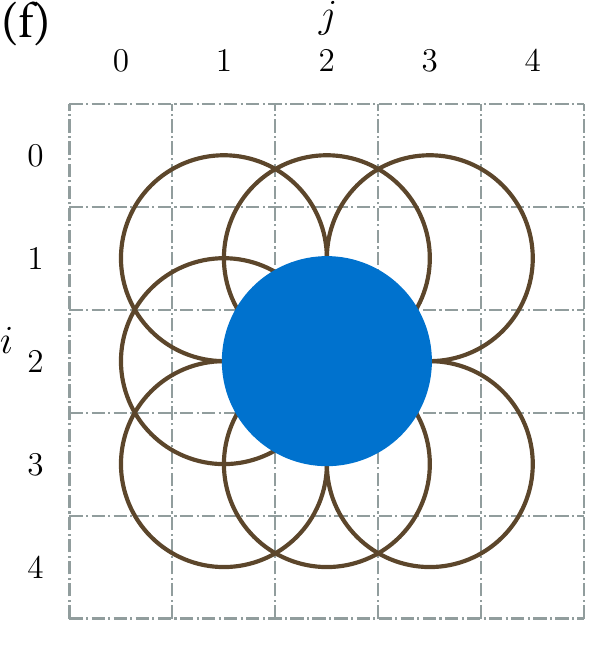}
        \caption{
                Besides drop patterns with 0 and 8 prior drops within the region of interest,
                6 more drop patterns are used to validate the model.
                From (a) to (f) are the drop patterns with 1 to 7 prior drops.
                The blue disk represents the new drop while the black circles
                represent the existing drop before the deposition.}
        \label{fig:all_patterns}
\end{figure}

Since this is an empirical model,
Sample size has to be sufficiently large.
Depending on the pattern of previously cured drops within \(B_{i j}\),
the new drop can spread differently.
We investigated 8 different patterns to validate that this model is not pattern dependent.
In addition to depositing cells with 0 and 8 surrounding cells,
other patterns are shown in \mbox{Figure~\ref{fig:all_patterns}}.
We prepared 75 samples for each pattern.
All 75 patterns are randomly separated in group of 3 for bootstrapping.
All coefficients are obtained from the first two passes using the pattern in Figure~\ref{fig:all_patterns}.(b)
with following equations
\begin{equation}
        \label{eq:coeff}
        \begin{aligned}
                m_v^+ & = \frac{\Delta v_{p q}[2]}{\tilde{h}_{p q}[1]} \hspace{1em} \tilde{h}_{p q}[1] > 0 \\
                m_v^- & = \frac{\Delta v_{p q}[2]}{\tilde{h}_{p q}[1]} \hspace{1em} \tilde{h}_{p q}[1] < 0 \\
                m_a^+ & = \frac{\Delta a_{p q}[2]}{\tilde{h}_{p q}[1]} \hspace{1em} \tilde{h}_{p q}[1] > 0 \\
                m_a^- & = \frac{\Delta a_{p q}[2]}{\tilde{h}_{p q}[1]} \hspace{1em} \tilde{h}_{p q}[1] < 0 \\
        \end{aligned}
        \hspace{1em} (p,q) \in B_{i j}.
\end{equation}

Using measured measured profile from printed patterns after the first and second pass,
the percent volume change and the percent area change can be obtained.
Using the drop pattern shown in Fig.~\ref{fig:all_patterns}.(b) as example,
the percent change in volume for neighboring cells around cell (2,2) after
a drop is deposited during the second pass of printing,
\(\Delta V_{2,2}[2]\) and the corresponding change in area \(\Delta A_{2,2}[2]\)
can be computed by:
\begin{equation}
        \label{eq:va_values}
        \begin{aligned}
                \Delta V_{2,2}[2] & = V_{2,2}[2] - V_{2,2}[1]                       \\
                                  & =   \left[\begin{array}{rrr}
                                1.65 & 8.96  & 3.93 \\
                                5.02 & 63.46 & 5.00 \\
                                2.34 & 7.94  & 1.71 \\
                        \end{array}\right] \%  \\
                \Delta A_{2,2}[2] & = A_{2,2}[2] - A_{2,2}[1]                       \\
                                  & =   \left[\begin{array}{rrr}
                                10.05 & 21.07                 & 9.94 \\
                                0     & \multicolumn{1}{c}{-} & 0    \\
                                10.09 & 18.59                 & 9.57 \\
                        \end{array}\right] \%. \\
        \end{aligned}
\end{equation}
Since the (2,2) cell of \(\Delta A_{i,j}[k]\) indicate the current drop location,
a drop on this location will cover the entire cell,
i.e. \(a_{2,2}[2] = 100\%\),
it is not calculated.
The height profile for the neighboring cells after the first pass,
\(H_{2,2}[2]\), can be obtained directly from measurement.
As a result, the height difference \(\tilde{H}_{2,2}[1]\) can be calculated with Eq.~(\ref{eq:h_d_s})
\begin{equation}
        \label{eq:h_value}
        \tilde{H}_{2,2}[1] = \left[\begin{array}{rrr}
                        -1.50 & -3.41 & -1.55 \\
                        6.24  & -1.03 & 6.12  \\
                        -1.42 & -3.77 & -1.40 \\
                \end{array}\right].
\end{equation}

Using experimental data shown in Eq.~(\ref{eq:va_values}) and (\ref{eq:h_value}),
The model coefficients can be calculated from Eq.~(\ref{eq:coeff}):
\begin{equation}
        \label{eq:coeff}
        \begin{array}{ll}
                m_v^+ = 0.0067 & m_v^- = -0.0201 \\
                m_a^+ = 0      & m_a^- = -0.0634 \\
        \end{array}.
\end{equation}

\section{EXPERIMENTAL RESULTS}
\label{sec:results}

Model validation is carried out with \(m_v^+\), \(m_v^-\), \(m_a^+\) and \(m_a^-\) from Eq.~(\ref{eq:coeff}).
The benchmark used to compare with other methods is root mean square (RMS) error of height within the ROI of each pattern,
\begin{equation}
        \centering
        \label{eq:rms}
        RMS = \sqrt{\frac{1}{9} \sum_{p=i-1}^{i+1} \sum _{q=j-1}^{j+1} \left (\frac{\bar{h}_{pq}[k]-h_{pq}[k]}{\bar{h}_{pq}[k]} \right)^2},
\end{equation}
where $\bar{h}_{pq}[k]$ represents the average height measured in the corresponding cell of all samples
and $h_{pq}[k]$ represents the predicted height in the corresponding cell after the $k^{th}$ pass is deposited at cell \((i,j)\).

\begin{figure}
        \centering
        \includegraphics[width=\linewidth]{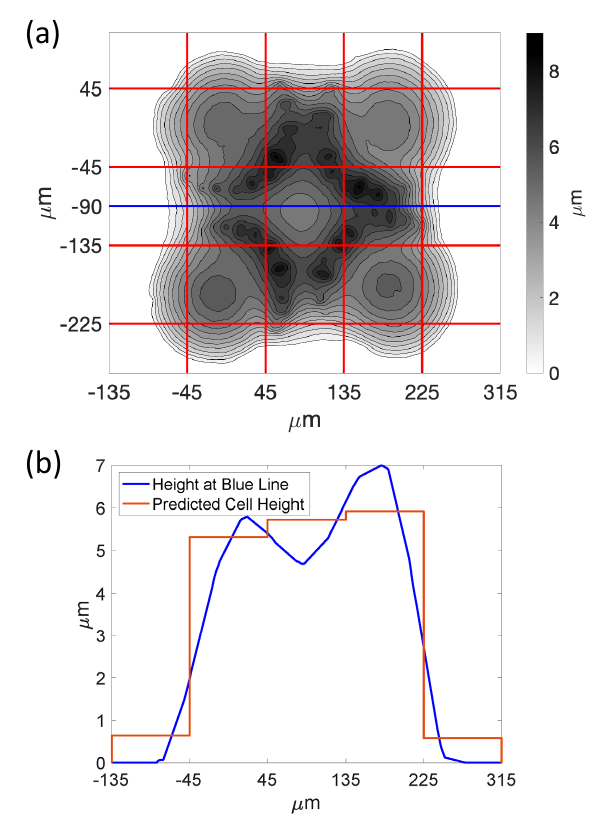}
        \caption{(a): One of the sample after a drop (3rd pass) is deposited at the center. Red lines are the cell boundary.
                The blue line indicates cross section measurements to be shown in (b).
                (b): The blue line shows the measured height along the blue line in (a). The red line shows the predicted cell height.}
        \label{fig:results}
\end{figure}

Figure~\ref{fig:results} shows the comparison between one of the sample and predicted height.
Figure~\ref{fig:results}.(a) is the contour plot of the measured height after the third pass,
during which a drop is deposited at the center.
All 8 surrounding cells have prior drops before the deposition.
The red lines mark the cell boundary, which are 90 \(\mu m\) apart.
The blue line marks the cross section along the center,
where the height measurements are plotted in Figure~\ref{fig:results}.(b) with the same color.
Figure~\ref{fig:results}.(b) compares the measurements with model predicted cell height,
where the red line represent the model predicted height.
The model predicted height follows the trend closely.

\begin{table}
        \centering
        \caption{RMS errors of model predicted cell height \\ (Patterns with different prior drops)}
        \label{tab:rms_error}
        \begin{tabular}{|c|c|c|c|c|c|}
                \hline
                \# of prior drops & 3      & 4      & 6      & 7      & 8      \\
                \hline
                RMS Error         & 4.54\% & 6.25\% & 3.35\% & 6.77\% & 5.84\% \\
                \hline
        \end{tabular}
\end{table}

\begin{table}
        \centering
        \caption{Comparison of RMS errors among different models \\ (1 prior drop)}
        \label{tab:rms_error_comp}
        \begin{tabular}{|c|c|c|}
                \hline
                Graph-based Dynamic & DSCC 2020 Model & This Model \\
                \hline
                17.2\%              & 5.9\%           & 5.11\%     \\
                \hline
        \end{tabular}
\end{table}

Table~\ref{tab:rms_error} shows the RMS errors of each pattern of different prior drops using Equation~\ref{eq:rms}.
The RMS errors of different patterns are consistently lower than 7\%,
validating that the model is not constrained to a specific pattern.
Using the same benchmark,
RMS errors of this model are compared with other two models.
The comparison of height prediction on 1 prior drop is shown in Table~\ref{tab:rms_error_comp}.
This method achieves similar RMS error level our previous 1D model. 
Moreover,
this method outperforms the reported graph-based dynamic model \cite{Guo2018}.

\section{\MakeUppercase{Conclusion}}
\label{sec:conclusion}
In this paper, a height profile model for 2D patterns for drop-on-demand printing of UV curable inks is proposed.
It follows the same approach as our previous height profile model for 1D patterns.
To ensure volume conservation,
both volume and area propagation are modeled as a function of height difference prior to the deposition.
To expand to 2D patterns,
height difference is calculated as the difference between the height of the cell and the average height of surrounding cells within
a region of interest. 
Based on experimental data,
the behavior of the ink flow differs by height differences
and the model is adjust accordingly.

The proposed model is validated experimentally with different 2D patterns.
The resulting RMS errors are consistently less than 7\% across different patterns.
The result is similar with our previous model for 1D patterns
and are better than the reported graph-base dynamic model.

\bibliographystyle{IEEEtran}
\bibliography{library.bib}
\end{document}